\documentclass[aps,eqsecnum,twocolumn,nofootinbib,prd]{revtex4}

\usepackage{graphicx}
\usepackage{amsmath}
\usepackage{amssymb}
\usepackage{amsbsy}
\usepackage{bm}

\begin{document}
\title{\Large On the scientific method learned from Albert Einstein in 2005---the World Year of Physics}
\author{Y. H. Yuan}
\email[E-mail: ]{henry@physics.wisc.edu}
\date{November 22, 2005}

\begin{abstract}

We review the physics at the end of the nineteenth century and 
summarize the process of the establishment of Special Relativity by 
Albert Einstein in brief. Following in the giant's footsteps, we outline 
the scientific method which helps to do research. We give some 
examples in illustration of this method. We discuss the 
origin of quantum physics and string theory in its early years 
of development. Discoveries of the neutrino and the correct model of 
solar system are also presented. 

\end{abstract}

\maketitle

\section{Historical review and a method deduced}

The year 2005 was declared the World Year of Physics which is an 
international celebration of physics. It marks the hundredth 
anniversary of the pioneering contributions of Albert Einstein, the 
greatest man in the twentieth century as chosen by Time magazine.

In 1905, one hundred year ago, a Swiss patent employee, Albert 
Einstein, published a paper entitled ``On the Electrodynamics of Moving 
Bodies" \cite{e1} which described what is now known as Special 
Relativity. It drastically changed human fundamental concepts of motion, 
space and time. The year 1905 was the miraculous year for Einstein. In the same year he also 
published three other trailblazing papers. One \cite {e2} accounted for 
the photoelectric phenomena and made up a part of the foundation of 
quantum mechanics. He won the Nobel Prize in physics due to the ideas 
of this paper in 1921. The second one \cite{e3} was about the 
explanation of Brownian motion and helped to establish the reality of 
the molecular nature of matter and to present convincing evidence for 
the physical existence of the atom. The third one \cite{e4} gave the 
most famous and beautiful equation in special relativity, $E=mc^2$, 
which has received various experimental verifications and has had wide 
application in modern physics. These groundbreaking papers have 
shattered many cherished scientific beliefs and greatly promoted the 
development of modern physics. They won for Einstein the greatest 
physicist as Newton in all human history. Next, we will follow the 
process of the establishment of Special Relativity and summarize some 
useful skills in research. 

One of the most famous puzzles at the end of the nineteenth century was the ether 
which was proposed as a medium to support the electromagnetic wave 
propagation. 
Maxwell's fundamental equations about the electromgnetic field were 
published in 1862. It leads to the electromagnetic wave equation 
in free space,
\begin{equation}
\nabla^2\phi -\frac{1}{c^2}\frac{\partial^2 \phi }{\partial{t^2}}=0
\end{equation}
where $\phi$ is any component of $\vec{E}$ or $\vec{B}$. Classical 
mechanics tells us that wave propagation needs a medium to support it. 
For example, sound waves have to travel in the air medium. 
For the propagation of electromagnetic wave, physicists presumed an 
ether medium which was entirely frictionless, pervaded all 
space, and was devoid of any interaction with matter. Although many 
ingenious physics papers during 1885-1905 were dedicated to verifying 
it, the ether refused to reveal its presence to the pursuers. 

In 1881, a 28-year-old American physicist, Albert Michelson, realised 
the possibility of an experimental test for the existence of ether 
by measuring the motion of the earth through it. He performed the 
experiment in Potsdam, Germany. Although he got 
negitive results in detecting the relative motion of the earth and 
ether, his measurement is not so accurate as to give an important 
result. Six years later, Albert Michelson and Edward Morley in 
Cleveland, Ohio carried out a high-precision experiment to demonstrate 
the existence of ether with an interferometer, which is now called 
the Michelson interferometer. This experiment is a high-precision 
repetition of Michelson's experiment in Potsdam. 
In their experiment shown in FIG. 1, a beam of light from the source was directed at an angle of 45 degree at a half-silvered mirror and was split into two beams 1 and 2 at point O by the mirror too. These beams 1 and 2 traveled at a right angle to each other. The two beams were 
reflected by separate mirrors, then recombined and entered a telescope to form a fringe 
pattern. The fringe pattern would shift if there was an effect due to 
the relative motion of the earth and the ether when the apparatus was 
rotated. Therefore, by monitoring the changes in the fringe pattern, 
they could tell the relative motion of the earth and the ether. Even 
with the high-precision 
apparatus, they did not find any experimental evidence for the 
existence of relative motion of the earth and ether. These results 
indicated that either there is no ether or the earth is in the ether 
rest frame all the time during the experiment. Since 
the earth is always altering its velocity when moving around the sun, 
the experimental result appeared to show that there was no existence of 
ether. All the repetitions of their experiment in the succeeding years 
were still unable to detect any relative motion of the earth and the 
ether. 

\begin{figure}[ht]
\includegraphics[width=\columnwidth,angle=0]{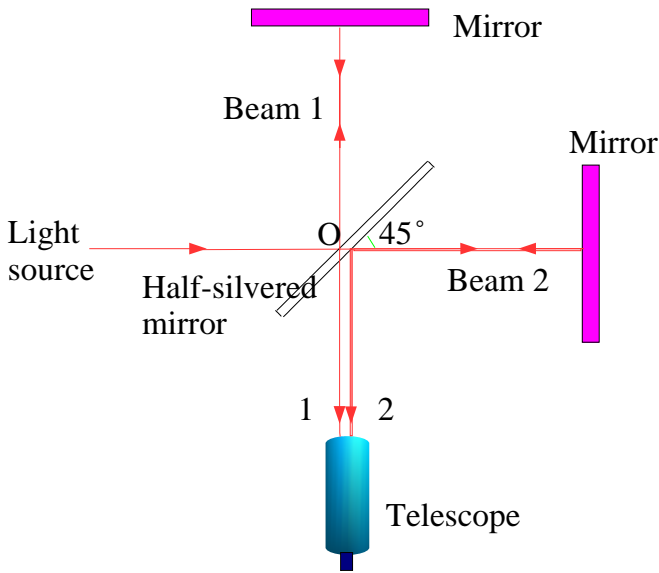}
\caption{Diagram of Michelson-Morley experiment. A beam from a source was split into beams 1 and 2 at point O by a  half-silvered mirror. The two beams were reflected by separate mirrors and then entered a telescope to form a fringe pattern. \label{f1}}
\end{figure}

But ether died hard. Many physicists including Michelson himself made 
great efforts to retain the ether yet explain the Michelson experiment. 
He attributed the negative results to the earth dragging some of the 
ether along with its motion. As a consequence the ether was 
motionless with respect to the earth near its surface. 

George FitzGerald put forward another possible explanation in 1892 
following the Lorentz-FitzGerald contraction equation, as we now 
know, 

\begin{equation}
L=L_0\sqrt {1-\frac{v^2}{c^2}}
\end{equation}
where $L_0 $ is called the proper length of an object which is measured 
in the rest frame of the object. To a moving observer of velocity $v$, 
any length along the direction of motion undergoes a length contraction 
by a factor of $\sqrt{1- \frac{v^2}{c^2}}$. He proposed that the 
experimental apparatus would shorten in the direction parallel to the 
motion through the ether. This shrinkage would compensate the light 
paths and prevent a displacement of the fringes due to the relative 
motion of the earth and the ether. 

Hendrik Lorentz discovered the well-known Lorentz transformation in 
1904 under which the electromagnetic theory expressed by the 
celebrated  Maxwell equations were in form invariant in all inertial 
frames.

\begin{eqnarray}
t'&=&\gamma (t-\frac{v}{c^2} x) \\
x'&=&\gamma (x-vt) \\ 
y'&=& y \\
z'&=& z
\end{eqnarray}
where we assume coordinate system $\Sigma$ moves relative to another 
one $\Sigma'$ in the $x$-axes with uniform velocity $v$ and 
$\gamma=\frac{1}{\sqrt {{1-\frac{v^2}{c^2}}}}$\cite{ja}.
Although he laid the foundation for the theory of relativity with his 
mathematical equations, Lorentz still tried to fit these 
remarkable equations into the ether hypothesis and save the ether from 
the contradiction of the Michelson experiment. 

All these efforts failed to explain the Michelson experiment while 
retaining the ether. It was genius Albert Einstein who abandoned 
the ether entirely. He wrote in his celebrated paper on relativity in 
1905\cite{an}: 

``The introduction of a `light ether' will prove to be superfluous, 
inasmuch as in accordance with the concept to be developed here, no 
`space at absolute rest' endowed with special properties will be 
introduced, nor will a velocity vector be assigned to a point of empty 
space at which electromagnetic processes are taking place." 

Furthermore he developed the Special Relativity by conjucturing two 
postulations:

1. The laws of nature are the same in all coordinate systems moving 
with uniform motion relative to one another. 

2. The speed of light is independent of the motion of its source.

Of which, the first one is a natural generalization for all kinds of 
physical experience since it is reasonable to expect that the laws of 
nature are the same with respect to different inertial frames of 
reference. Whereas the second one just represents a simple experimental 
fact. In Michelson's experiment, the speed of light was found to be 
constant with respect to the earth. Put in other words, the speed of 
light is the same for observers in different inertial frames of 
reference since an observer on the earth at two different times may be 
regarded as an observer in two different inertial frames of reference. 
It is only a small jump to the postulate of Einstein's Special Relativity that the speed 
of light is independent of the motion of its source. 

From the above condensed outline of the establishment of special 
relativity, we learn that when proposing an idea or theory with which we may 
account for some unexplained phenomena, they should be based upon the 
given facts of experiment or phenomena. Sometimes, the ideas may 
contradict well-known theories which are not experimentally proved, 
such as the ether. If the problem is subtle and complicated,
especially in physics, we should make incisive analyses, see through 
the general appearance and grasp the underlying nature. The above method is of practical
use in research which could be seen from the following four examples.

\section{Four examples}

A good example in illustration of the method is the origin of quantum physics which now plays an 
important role in various scientific areas. At the end of the nineteenth 
century, classical physics achieved great success. But some 
experimental results were incompatible with the classical physics such 
as the specific heat of a solid, the photoelectric effect and the 
thermal radiation of a black body. Kirchhoff initiated his theoretical 
research on thermal radiation in the 1850s. By the end of the 
nineteenth century two important empirical formulae on black-body 
radiation had been derived based upon the fundamental thermodynamics. 
Wien proposed a formula for the energy density inside a black body in 1896,

\begin{equation}
\rho(\nu,T)= c_1 \nu^3 e^\frac{c_2 \nu}{T}
\end{equation}
where $T $ is the temperature of the wall of a black body, $\nu$ is the 
radiation frequency and $c_1$,$c_2$ are 
two constants. Rayleigh and Jeans derived a result from a different 
approach in 1990,

\begin{equation}
\rho ( \nu,T)= \frac{8\pi \nu^2}{c^3} k T
\end{equation} where $k$ is the Boltzmann's constant. The Rayleigh-Jeans formula was in agreement with the experimental curve 
at low frequencies whereas the Wien formula fitted the experimental 
curve well at high frequencies. It was a great discrepancy at the turn 
of twentieth century that they failed to completely explain black-body 
radiation. Seemingly there was no way out since these formulae were 
based upon the fundamentals of classical physics. 

Things changed on Dec. 14, 1900 when at a German Physical Society 
meeting Max Planck presented his paper entitled ``On the theory of 
energy distribution law of normal spectrum" which not only solved the 
puzzel of black-body radiation but uncovered the quantum world. It 
marks the birth of quantum physics. He assumed that this energy could 
take on only a certain discrete set of values such as 0, $h\nu, 2h\nu, 
3h\nu,$..., where $h$ is now known as Planck's constant. These values 
are equally spaced rather than being continuous. This assumption 
apparently contradicted the equipartition law and common sense. He 
argued that the wall of a black body emitted radiation in the form of 
quanta with energy of integer mutiple of $h \nu$. Based on this bold 
assumption\cite{mp}, Planck gave a formula of the energy density at frequency 
$\nu$,

\begin{equation}
\rho ( \nu,T)= \frac{8\pi\nu^2}{c^3} \frac {h \nu}{e^\frac{h\nu}{kT}-1}
\end{equation}
which were in complete agreement with experimental results on general 
grounds. His formula was an ingenious interpolation between 
the Wien formula and the Rayleigh-Jeans formula. We now know that it 
gives the correct explanation of the black body radiation 
spectrum. This proposal established his status in science. As Einstein said\cite{pl}:

``Very few will remain in the shrine of science, if we eliminate those moved by ambition, calculation, 
of whatever personal motivations; one of them will be Max Planck.''
 
Another example concerns the situation of string theory in its early 
stage. We know that quantum field theory worked well in the 
unification of quantum mechanics and electromagnetism in the 1940s. 
That it could also describe the weak and strong interactions 
was understood by the end of 1960s. It has played a significant role in 
our understanding of particle physics in many ways, from 
the formulation of the four-fermion interaction theory to the 
unification of electromagnetic and weak interactions
\cite{gl}\cite{we}\cite{sa}. But when we attempted to incorporate it 
with gravity at high energy scale, severe problems appeared. 
For example, when $E>M_p$, the interaction of gravitation can not be 
negligible, where 
\begin{equation}
M_p=(\frac{{\hbar}{c}}{G})^{1/2}{c^2}\approx 1.2\times 10^{19} Gev
\end{equation}
is Planck energy. The short-distance divergence problem of quantum 
gravity arouse. It was non-renormalizable even though we have employed the 
usual renormalization extracting the meaningful physical terms from the 
divergences. 

String theory solved this severe problem. According to the postulates in string theory, all 
elementary particles, as well as the graviton were regarded as one 
dimensional strings rather than point-like particles which 
were generally accepted at the time. But the generally accepted 
point-like particle concept was not experimentally proved.
Each string has a lot of different harmonics and the different elementary particles were regarded as different harmonics in string theory. Therefore the world-line of a particle in quantum field theory shown in FIG. 2 was replaced by its analog in string theory, the world-sheet of a string which could join the world-sheet of another string smoothly. As a consequence, the vertex of an interaction in a Feynman diagram was smeared out. In string theory the massless spin two 
particle in the string spectrum was just right identified as the 
graviton which mediates gravitation. At low energy scale the 
interaction of massless spin two particle is the same as that required by general 
relativity. From this simple string postulate, string theory leads to a number of 
fruitful results. It is the only currently known consistent theory 
of quantum gravity which does not have the above divergence problem. 
One of the vibrational forms of the string possesses just the 
right property---spin two---to be a graviton whose couplings at long distance are those of general relativity. It admits chiral gauge 
couplings which have been the great difficulty for other 
unifying models. In addition, string theory predicts supersymmetry and generates Yang-mills gauge fields and it has found many applications to mathematics in the area of topology and geometry. Also in string theory there are no dimentionless adjustable parameters which generally appear in quantum field theory, such as the fine-structure constant $\frac{e^2}{4\pi\epsilon_0 \hbar c}$.
Nowadays, string theory (Detailed descriptions may be found in Refs.\cite{gsw}) has already become one of the most active areas of 
research in physics. Thanks to the postulate of one dimensional string. It sheds light on and promises new insights to some 
deepest unsolved problems in physics, for example, what cause 
the cosmic inflation? How does time begin? What constitutes the dark matter and what is the so-called dark energy?

The third example is upon the process of the discovery of the neutrino.(In this paper, all neutrino mean $\nu_e$ only.) In 1896, 
radioactivity was discovered by Henri Becquerel 
which marked the birth of modern nuclear physics. Subsequently three types of radioactive rays were identified. They were called alpha ray, beta ray and gamma ray separately.
Becquerel established that the beta rays were high-speed electrons in 1990.
Employing electric and magnetic fields, he deflected beta rays 
and found that they were negatively charged and that the ratio of charge 
to mass of the beta particle was the same as that of an electron. After more accurate
measurements on beta decays physicists found a serious problem. Unlike 
alpha decay and gamma decay in which the emitted particles carried away
the well-defined energy which is equal to the total energy difference of the 
initial and final states, beta decay emited electrons with a
continuous energy spectrum. It meant that a particular nucleus emitted an 
electron bearing unpredictable energy in a particular transition. This
experimental result apparently violated the conservation laws of energy and momentum. Wolfgang Pauli proposed an entirely new particle-neutrino in order to solve this serious problem. In his open letter\cite{pa} to the group of radioactives at the meeting 
of the regional society in Tubingen on December 4, 1930, he proposed the neutrino based on the given fact of experiment:

``...This is the possibility that there might exist in the nuclei electrically 
neutral particles, which I shall call 
neutrons, which have spin 1/2, obey the exclusion principle and 
moreover differ from light quanta in not travelling with the velocity 
of light.''

``... I admit that my remedy may perhaps appear unlikely from 
the start, since one probably would long ago have seen the neutrons if 
they existed. But `nothing venture, nothing win', and the gravity of 
the situation with regard to the continuous beta spectrum is 
illuminated by a pronouncement of my respected predecessor in office, 
Herr Debye, who recently said to me in Brussels `Oh, it's best not to 
think about it at all-like the new taxes'. One ought to discuss seriously every avenue of rescue."

In his letter, Pauli called his new proposed particle-the "neutron" which is now called neutrino due to Enrico Fermi. Pauli proposed that this new speculative neutral particle might resolve the nonconservation of energy. If the proposed neutrino and the electron were emitted simultaneously, the continuous spectum of energy 
might be explained by the sharing of energy and momentum of emitted 
particles in beta decay. It is worth mentioning that long before the neutrino was experimentally detected, Enrico Fermi\cite{fe} incorporated Pauli's proposal in his brilliant model for beta decay in the framework of quantum electrodynamics in 1934. He showed clearly with his beta decay theory that the neutron decayed into a proton, an electron and a neutrino simultaneously. The neutrino was experimentally detected by Fred Reines and Clyde Cowan, at the Los Alamos lab in 1956\cite{re} using a liquid scintillation device. This important discovery won the 1995 Nobel prize in physics.  A lot of famous phenomena and problems solved and unsolved, related with the neutrino were found. Parity violation takes place whenever there is the neutrino taking part in a weak interaction. This is just as the behavior of monopole under parity.
They may be the same particle we think drawing inspiration from Einstein. So it causes the above violation. Time reversal violation of this kind of weak interaction also is due to sign change of charges. When detecting neutrino emitted from the sun, fewer solar neutrino capture rate than the predicted capture rate in chlorine from detailed models of the solar interior was found in 1968. This is the solar neutrino puzzel. We explain the flavor change easily by the new nature of neutrino in solar neutrino puzzel. In short, the change is by pair creation and annihilation.
Later the same phenomena were also observed by other groups using different materials. As the most fascinating particle, the neutrino is so important that neutrino physics has become one of the most significant branches of modern physics. Thanks to the conjecture of the neutrino by Pauli. Although his proposal contradicted the well-accepted knowledge at the time on beta decay process, his new beta decay process involving the neutrino was not completely impossible experimentally. With this proposal we could overcome the serious problem and rescue the fundamental conservation laws of energy and momentum. The little neutrino has found its application to a number of different research areas in physics, such as in particle physics, nuclear physics, cosmology and astrophysics.

\begin{figure}[ht]
\includegraphics[width=\columnwidth,angle=0]{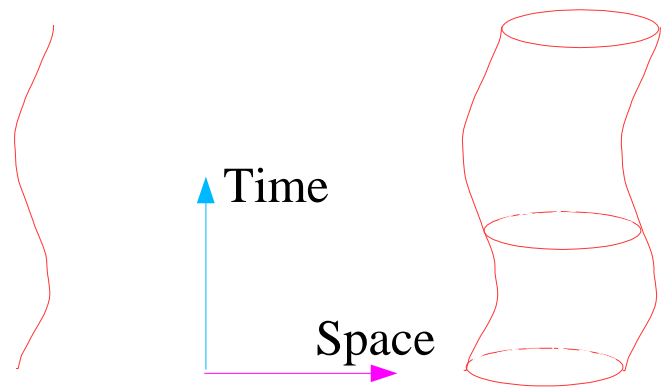}
\caption{A point particle's world-line (left) in spacetime and its analog in string theory, world-sheet, traced out by a closed string (right) in spacetime. \label{f2}}
\end{figure}

Finally we give an example of astronomy to show the usefulness of 
our method deduced from the establishment of special relativity.
Before the sixteenth century, it was extensively accepted that the sun, the moon and planets all orbited about the earth which was at the center of the universe.
In his famous book, Almagest, the antient Greek astronomer Claudius Ptolemy proposed the earth-centered model of the universe. He proved that the earth was 
round and the gravity everywhere pointed to the center of the earth. 
Every planet moved along an epicycle whose center revolved around the earth
just as the sun and the moon revolving around the earth. The epicycle was carried along on a larger circle like a frisbee spinning on the rim of a rotating wheel shown in FIG. 3. He postulated the 
epicycle to explain the looping motion of a planet. The people on the 
earth would not see the observed irregular motion of a planet if the 
planet moved around the earth in a circular orbit, rather than in a epicycle. The Ptolemaic model of the universe was the obvious and 
direct inference when people observed the motion of the sun day after day and the motions of the moon and the planets night after night. Therefore Ptolemy's theory was well-accepted and prevailed for a long time. 

\begin{figure}[ht]
\includegraphics[width=\columnwidth,angle=0]{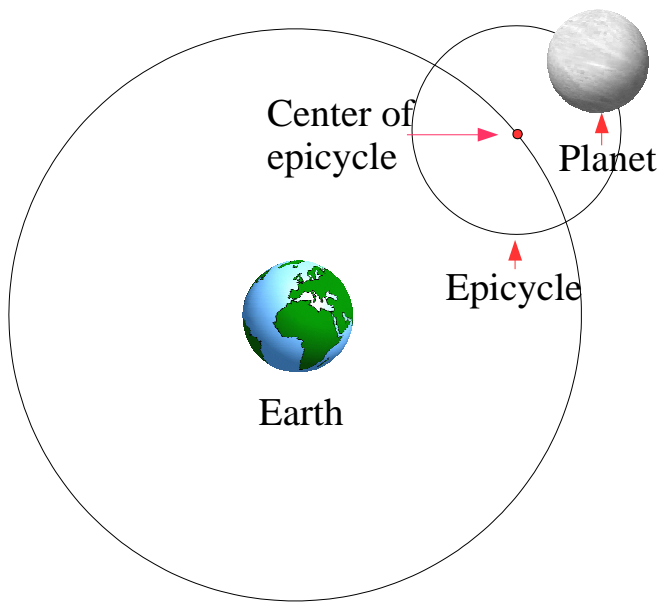}
\caption{An illustration of the Ptolemaic model. \label{f4}}
\end{figure}

In about 1510, 
Nicholas Copernicus presented the helio-centric model. In his 
celebrated book published in 1543, De Revolutionibus orbium coelestium, 
he postulated that the planets including the earth all moved around the 
sun shown in FIG. 4\cite{nc}. (In 1781, William and Caroline Herschel discovered Uranus, the first planet found beyond the Saturn boundary, which was generally acknowledged as the outer limit of the solar system for thousands of years. Neptune was discovered in 1846 by Johann Galle. The discovery of Neptune was a great triumph for theoretical astronomy since Neptune was at first predicted by Adams and Le Verrier using mathematical arguments based on Newton's universal gravitation law and then observed near their predicted locations. Pluto was predicted by Percival Lowell and found in 1930 by Clyde Tombaugh.) The earth spin about its axis one rotation per day and revolved around the sun in the plane of the ecliptic. He explained the apparent 
looping motions of the planets in a simple way using his new helio-centric model. They were the direct consequence of the relative motion of the 
planets and the earth when people saw from the earth.

He could not prove his radical helio-centric model at the time. Although he simplified the cumbersome Ptolemaic system, both 
the earth-centered model and the helio-centric model could account for the observations of motion of the celestial bodies. Copernicus's theory gives an alternative 
theory of Ptolemy. Even though the problem was subtle and complicated, we should 
grasped the hidden nature behind the phenomena. As Copernicus pointed out that 
the extremely massive sun must rule over the much smaller planet and the earth. He therefore drew 
his conclusion that it was the earth that moved around rather than the 
sun. It was Isaac Newton who provided the correct explanation of Kepler's laws and 
convinced people that the earth and other planets revolved around the 
massive sun due to the attractive force with his ingenious universal 
law of gravitation\cite{nt}. 

\begin{equation}
F=G\frac{Mm}{r^2}
\end{equation}

where $F$ is the universal gravitation force between the two bodies 
with mass M and m respectively, G is called the gravitational constant, and r is the 
distance between the centers of mass of the two bodies. Galileo Galilei first observed the four satellites orbited around Jupiter which exhibited undoubtedly that the earth were not the center of all circular motions of the celestial bodies utilizing his telescopes. He stated the four small bodies moved around the 
larger planet-Jupiter like Venus and Mercury around the sun. Also he observed the phases of 
Venus which was the direct result of the planet moving around the sun.

\begin{figure}[ht]
\includegraphics[width=\columnwidth,angle=0]{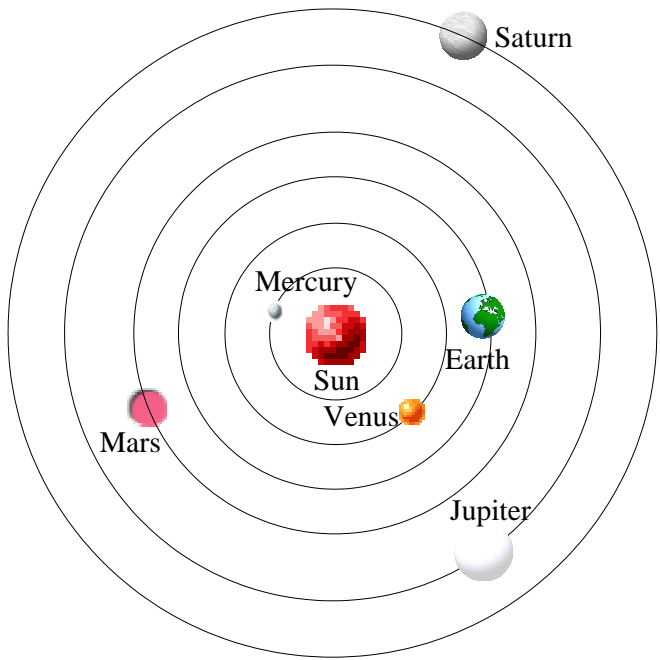}
\caption{An illustration of the Copernican model in which the planets, including the earth, all orbited around the 
sun. His heliocentric concept 
contradicted Ptolemy's model, the well-accepted model treated with respected at the time. \label{f3}}
\end{figure}

\section{Summary and conclusion}

We have reviewed the process of the establishment of Special Relativity 
against the background of physics around the turn of the twentieth century 
in our paper. Moreover we have outlined the scientific method which helps to do 
research. Some examples are presented in order to illustrate the 
usefulness of the method. We have discussed the origin of quantum 
physics and string theory in its early years of development. 
Discoveries of the neutrino and the correct model of solar system have also been demonstrated. We have shown that the method is of practical use in a wide range, from physics to astronomy, from ancient science to modern ones. This year is the unprecedented World Year of Physics which 
acknowledges the contribution of physics to the world. It marks the 
hundredth anniversary of the pioneering contributions of Albert 
Einstein in 1905 as well as the fiftieth anniversary of his death in 
1955. We dedicate this paper to Albert Einstein.

\end{document}